\documentclass[aps,prl,twocolumn,superscriptaddress, amsmath,amssymb]{revtex4-2}

\usepackage{graphicx}
\usepackage[linkcolor=blue]{hyperref}
\usepackage{dcolumn}
\usepackage{bm}
\usepackage{xcolor}
\begin{document}


\title{Quantifying Robustness of Attosecond Transient Absorption Spectroscopy for Vibronic Coherence in Charge Migration}


\author{C. H.  Yuen}
\email[]{iyuen@phys.ksu.edu}
\author{C. D.  Lin}
\email[]{cdlin@phys.ksu.edu}
\affiliation{J. R. Macdonald Laboratory, Department of Physics, Kansas State University, Manhattan, Kansas 66506, USA}

\begin{abstract}
Probing vibronic coherence in molecules has been a central topic in ultrafast science,  as it is an essential prerequisite to monitoring electronic motion.
While experiments have demonstrated that attosecond transient absorption spectroscopy (ATAS) can probe the vibronic coherence in a few molecules,  
its robustness remains largely unexplored.
In this Letter,  we develop a comprehensive theory for ATAS which accounts for the orientation dependence of the density matrix of a pumped molecule.
We apply our theory to N$_2^+$ formed by multiorbital tunnel ionization under a few-cycle intense near-infrared laser pulse.
The simulated x-ray absorption spectrum shows clear signatures of the vibronic coherence between the $A^2\Pi_u$ and $B^2\Sigma_u^+$ states of N$_2^+$,  
which was predicted to be absent by a previous theory.
We further define a coherence contrast factor to quantify the robustness of ATAS.
This work advances the theoretical foundation of ATAS and paves the way for monitoring electronic motion in generic molecules.
\end{abstract}


\maketitle


The capability of monitoring coherence in molecules is the next milestone in ultrafast science.  
This is because coherence drives the movement of electron density,  which is known as charge migration.
Charge migration can be triggered by creating a coherent superposition of electronic states through laser excitation or ionization~\cite{cederbaum1999, remacle2006,  folorunso2021}.
But within a few femtoseconds (fs),  nuclei of the molecular ion start moving,  and the coherence between different electronic states could undergo rapid changes and could even vanish,  leading to permanent charge transfer.
As a result,  monitoring the change of coherence not only allows real-time observation of electronic motion,
but could also lead to the control of the ion's chemical reactivity~\cite{remacle1998,  lepine2014},
which are the ultimate goals of ultrafast science~\cite{biegert2021}.

While charge migration or similar experiments have been conducted for various molecules using different techniques over the years~\cite{smirnova2009,sansone2010, calegari2014,  kraus2015,  okino2015,  kobayashi2020a,kobayashi2020,  schwickert2022, He2022, matselyukh2022},  there are still two major challenges to reach the goals.
The first challenge is in the time window for measurement.
To observe the change of electronic coherence due to nuclear motion,  which we refer to as vibronic coherence,  the time window should be in tens of fs.
This requirement restricts the use of high-harmonic spectroscopy,  which has been used to probe charge migration in several molecules~\cite{smirnova2009,kraus2015,He2022},  due to its limited time window within an optical cycle.
This limitation arises from its mechanism,  known as the three-step model~\cite{corkum1993},  where an electron undergoes tunnel ionization followed by acceleration in the driving laser field.
Within the optical cycle of the laser (a few fs),  it recombines with the molecular ion and emits high harmonics photons carrying information about the coherence.
Therefore,  to extend the time window,  two laser pulses should be used in the experiments,  in which the first pulse triggers the charge migration and the second pulse generates the observables at the relevant time window.
This leads to the second challenge: the observables must be sensitive to the change of coherence and the probing mechanism should be well understood.
However,  it has been complicated to simulate most pump-probe schemes employed in experiments~\cite{calegari2014,  okino2015,  schwickert2022,  matselyukh2022}; either because too many pumped states were involved or the probing mechanism is either unknown or intractable theoretically.
This leaves an important question: Are there robust pump-probe schemes to study vibronic coherence in charge migration?

So far,  two promising pump-probe schemes for vibronic coherence have been simulated: dissociative dication momentum spectroscopy (DDMS)~\cite{yuen2023b} and attosecond transient absorption spectroscopy (ATAS)~\cite{kobayashi2020a,  golubev2021,  kobayashi2022t}.
In DDMS,  the neutral molecule is first ionized by either a few-cycle infrared (IR) pulse or a short extreme ultraviolet (XUV) pulse to form a superposition of ionic states.
At a later time,  a few-cycle intense IR pulse further ionizes the pumped states to form dissociative dications.
Since this process is partly driven by laser couplings~\cite{yuen2022,  yuen2023},  changes in vibronic coherence between the pumped states lead to changes in the dication yields,  thereby imprinting the coherence to observables such as the kinetic energy release spectrum. 
Although there is remarkable theoretical evidence for this pump-probe scheme~\cite{yuen2023b},  experimental verification is currently in progress.
On the other hand,  in ATAS,  the neutral target is typically pumped by a few-cycle IR pulse,  then probed using the absorption of an attosecond XUV or x-ray pulse.
ATAS is widely considered a superior technique since it has excellent time resolution and can selectively probe the atoms in a molecule.
It has been shown that ATAS can probe the electronic coherence in atomic ions~\cite{goulielmakis2010} and vibronic coherence in halogen-containing molecular ions~\cite{kobayashi2020a,kobayashi2020}.
The decoherence and revival in charge migration in the excited silane molecule were also observed using ATAS~\cite{matselyukh2022},  making it a promising pump-probe scheme for charge migration.

For ATAS,  however,  there is an oversimplification in the theoretical treatment for molecules: Molecular orientation has not been considered.
The existing theory for molecules~\cite{kobayashi2020a,  golubev2021,  kobayashi2022t} have been directly derived from the theory for atoms~\cite{santra2011},  where the density matrix of aligned molecules simply replaces that of atoms.
But laser interaction with atoms and molecules differs significantly due to the dependence of laser-molecule interaction on molecular orientation.
As a consequence,  after interacting with the pump pulse,  population and coherence of the pumped states are anisotropic.
The oversimplification made in the existing theories results in isotropic population and coherence of the pumped states,  implying that only the coherence between states with similar symmetry can be detected~\cite{golubev2021,  kobayashi2022t}.
This implication severely undermines the robustness of ATAS,  considering that charge migration in molecules is typically triggered by superposition of electronic states with different symmetries.

In this Letter,  we address this key issue by considering the anisotropy of population and coherence of the pumped states and re-derive the ATAS theory for molecules.
By applying this theory to N$_2$,  we demonstrate the capability to probe coherence between the $A^2\Pi_u$ and $B^2\Sigma_u^+$ states of N$_2^+$ using ATAS.
To quantify the robustness of ATAS for probing charge migration in generic molecules,  we also introduce a coherence contrast factor that crucially relies on orientation dependence of density matrix.

To obtain an orientation-dependent density matrix for pumped molecules,  we model the tunnel ionization of the neutral and post-ionization dynamics of the ion under a few-cycle IR pulse using the density matrix approach described in Ref.~\cite{yuen2023c}.
The equation of motion for density matrices $\rho^{(q)}$ ($q=0,  1$ for neutral or ionic states) is
\begin{align}
 \frac{d}{dt}\rho^{(q)}(t) = -\frac{i}{\hbar}[H^{(q)}(t),\rho^{(q)}(t)] + \Gamma^{(q)}(t),
 \label{eq:EOM}
\end{align}
with the ionization matrix
\begin{align*}
 \Gamma^{(0)}(t) &= -\sum_i \rho^{(0)}(t) W^{(0)}_i,  \\
 \Gamma^{(1)}_{ij}(t) &= \rho^{(0)}(t) \sum_m \gamma^{(0)}_{im} (t) \left(\gamma^{(0)}_{jm}(t)\right)^\ast.
\end{align*}
In the above,  $H^{(q)}$ is the Hamiltonian with the laser coupling term $-\vec{d}\cdot \vec{E}$.
$W^{(0)}_i$ and $\gamma^{(0)}_{im} $ is the molecular Ammosov-Delone-Krainov ionization rate and amplitude from the neutral to the $i$th ionic state,  respectively~\cite{tong2002, yuen2023c}.
By using the ionization amplitude in $ \Gamma^{(1)}_{ij}$,  the coherence arise from tunnel ionization is taken into account~\cite{yuen2023c}.
Solving the above equation then yields density matrices of the neutral and the ion at different orientations.
The full density $\rho$ can be constructed by assuming the density matrix of different charge states is block-diagonal.

During the time delay between the pump and probe pulses,  the nuclear motion of the ion sets in.
We assume each vibronic state is populated according to the Franck-Condon (FC) factor $|c_{iv}|^2$ from the neutral vibronic ground state.
Then,  the nuclear wave packet for the $i$th state evolves as $|\chi_i (t)\rangle = \sum_v |c_{iv}|^2 e^{-i E_{iv}t} |\phi_{iv} \rangle$,
with $|\phi_{iv} \rangle$ and $E_{iv}$ being the vibronic wave function and energy.
The evolution of vibronic coherence between ionic states during the pump-probe delay,  under the FC approximation~\cite{yuen2023b},  is then 
\begin{align}
\rho^{(1)}_{ij}(t) = C_{ij} (\theta) \langle \chi_j (t-t_1) | \chi_i (t-t_1) \rangle,
\label{eq:FCF}
\end{align}
where $C_{ij} (\theta)$ matches the density matrix when the pump pulse ends ($t=t_1$).

For the probe pulse,  we assume that it is an attosecond pulse with a parallel polarization as the pump pulse.
The experimental observable here is the optical density (OD),  which is proportional to the absorption cross section $\sigma^{(1)} (\omega)$, 
\begin{align*}
\sigma^{(1)} (\omega) &= 4 \pi \frac{\omega}{c} \mathrm{Im}[\chi^{(1)} (\omega)],
\end{align*}
with $\omega$ being the probe frequency and $\chi^{(1)}$ being the linear susceptibility.
In Sec.  S1 of the supplementary material (SM),  we rigorously derived the expression for the orientation averaged linear susceptibility,
\begin{align}
\bar{\chi}^{(1)}(\omega,  \Delta t)
&= \sum_{ij}   \sum_f  \sum_{\mu, \nu} \frac{d^\ast_{fj,  \nu}  d_{fi,  \mu} \lambda_{ij;\nu \mu} (\Delta t)}{E_f - E_i  -\omega - i\epsilon} ,  
\label{eq:chi} \\
\lambda_{ij;\nu \mu} (\Delta t) &=  \frac{1}{4\pi}  \int_0^{2\pi} \int_0^{\pi} \rho_{ij}  (\Delta t,  \alpha,  \beta) \left[D^1_{\nu 0} (\alpha,  \beta,  0)\right]^\ast \nonumber \\
& D^1_{\mu 0} (\alpha,  \beta,  0)  d \alpha \sin\beta d\beta,
\label{eq:lambda}
\end{align}
where $D$ is the Wigner $D$-matrix and $(\alpha,  \beta)$ are the Euler angles which resemble the azimuth angle and polar angle with respect to the laser polarization on the $z$-axis.
The density matrix $\rho_{ij}$ depends on the pump-probe delay $\Delta t$ as well as the Euler angles.
$d_{fi,  \mu}$ is the core-valence transition dipole moment (TDM) of different spherical components $\mu$ in the molecular frame.
$E_f$ and $E_i$ are the electronic energy of the final core hole state $f$ and the valence state $i$, 
and $\epsilon$ accounts for the width of absorption spectra in experiments.
If the density matrix is isotropic,  Eq.~\eqref{eq:chi} reduced to 
\begin{align}
\bar{\chi}^{(1)}(\omega,  \Delta t) &= \frac{1}{3} \sum_{ij}  \rho_{ij}  (\Delta t) \sum_f  \frac{  \sum_{\mu} d^\ast_{fj,  \mu}  d_{fi,  \mu}} {E_f - E_i  -\omega - i\epsilon},
\label{eq:chi_wrong}
\end{align}
which is identical to the expression in Ref.~\cite{golubev2021} without the negative frequency term. 
This implies that coherence between state $i$ and $j$ can contribute to the overall signal only if the dot product of their TDM to state $f$ is non-zero.
Consequently,  Eq.~\eqref{eq:chi_wrong} suggests that signal from coherence between a $\Sigma$ and a $\Pi$ state of a linear molecule vanishes,  as their TDMs to state $f$ must be perpendicular.
This is in sharp contrast with the prediction from Eq.~\eqref{eq:chi},  in which such coherence signal is non-vanishing due to the anisotropy of density matrix.

An important observation is that,  in Eq.~\eqref{eq:chi},  coherence between state $i$ and $j$ can be probed only if both states can reach the same final state $f$.
It implies that only coherence between electronic states with the same parity can be probed for molecules possessing inversion symmetry.
It is opposite to the case of using the DDMS to probe vibronic coherence~\cite{yuen2023b}.
Since its probing mechanism relies on laser coupling between state $i$ and $j$,  coherence signal is strong only if the two states have the opposite parity.
As a result,  ATAS and DDMS could be used complementarily to fully characterize vibronic coherence in molecules with inversion symmetry.

\begin{figure}[ht]
\includegraphics[scale=0.5]{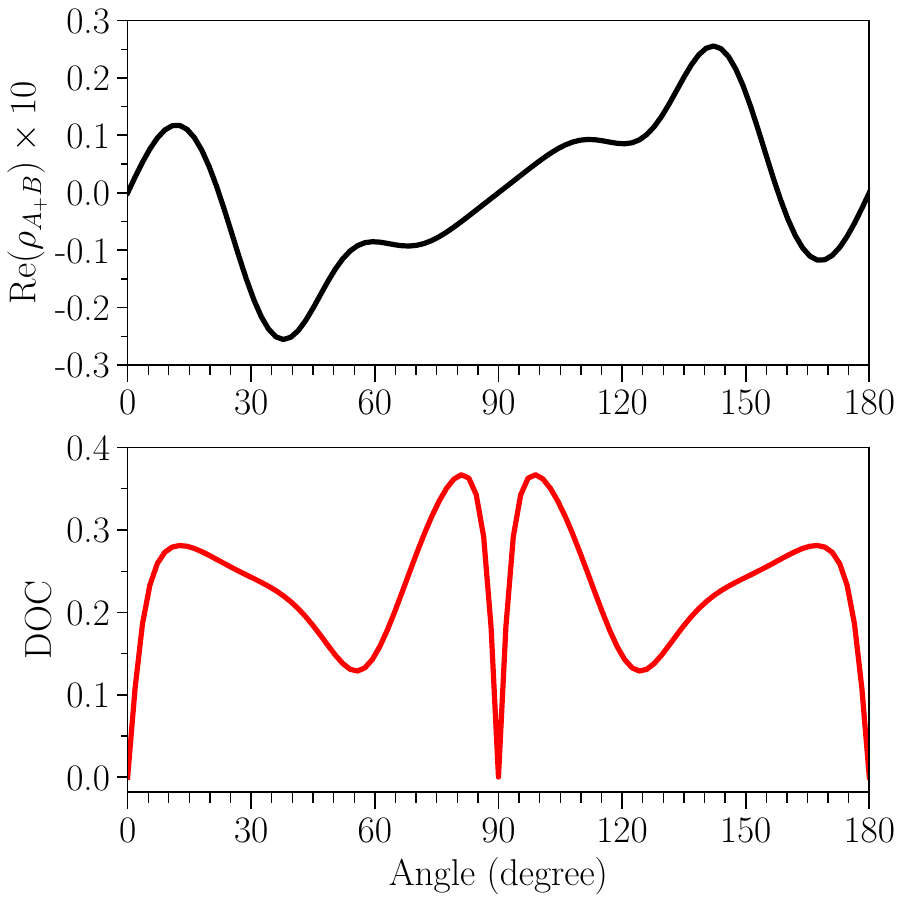}
\caption{Orientation dependence of the real part of the density matrix elements $\rho_{A_+B}$ of N$_2^+$ (top) and its degree of coherence (bottom) after the pump pulse.}
\label{fig:rhoab}
\end{figure}

We now apply our ATAS theory to N$_2$.
The pump pulse is assumed to be a 6 fs,  800 nm,  linearly polarized Gaussian pulse with a peak intensity of 3 $\times 10^{14}$ W/cm$^2$.
The pump pulse populates the $X^2\Sigma_g^+$ ($3\sigma_g^{-1}$,  15.6 eV),  $A^2\Pi_{u \pm}$ ($1\pi_u^{-1}$,  16.9 eV),   and $B^2\Sigma_u^+$ ($2\sigma_u^{-1}$,  18.8 eV) states of N$_2^+$ from the $X^1 \Sigma_g^+ (\ldots 2\sigma_u^2 1\pi_u^4 3\sigma_g^2)$ state of N$_2$.
For brevity in the discussion,  we refer to the neutral state as $\tilde{X}$ and the ionic states as the $X$,  $A_\pm$,  and $B$ states hereafter.
After the tunnel ionization,  the $X$ and $A_\pm$ states and the $X$ and $B$ states are coupled by the pump pulse.
Evolution of population and coherence in the ion is then further driven by the laser couplings.
By solving Eq.~\eqref{eq:EOM},  we obtain the density matrix of the neutral and the ion after the pump pulse at each molecular orientation.
During the pump-probe delay,  the vibronic coherence in the ion evolves according to Eq.~\eqref{eq:FCF},  with the FC factors and vibronic energies given by Ref.~\cite{gilmore1992}. 
The population of the neutral and the ion are constant throughout.
On the other hand,  to simulate the OD from the probe pulse,  we have included the following absorption lines: from $\tilde{X}$ to $1\sigma_u^{-1} 1\pi_g^{1}$ (401 eV),  from $X$ to $1\sigma_u^{-1} 3\sigma_g^{-1} 1\pi_g^{1}$ (402,  403 eV),  from $X$ to $1\sigma_u^{-1}$ (393.5 eV),  from $A$ to $1\sigma_g^{-1}$(392 eV),  and from $B$ to $1\sigma_g^{-1}$ (391 eV) (see Sec.  S2 in the SM for details).
The energies and TDMs for the transitions are adopted from a similar ATAS experiment for N$_2$~\cite{Kleine2022} (and Refs.~\cite{glans1996,  lindblad2020}),  in which pulse duration of the IR and soft x-ray pulse is 50 and 25 fs,  much longer than what we considered.
Note that Eq.~\eqref{eq:chi} does not depend on the soft x-ray pulse profile,   as long as the short pulse approximation $E(t) \sim \delta(t)$ is valid and the spectral range covers the relevant absorption lines.

As ATAS can only probe the coherence between the $A_\pm$ and $B$ states (same parity),  density matrix element $\rho_{A_\pm B}$ is of particular interest.
The top panel in Fig.~\ref{fig:rhoab} shows the real part of $\rho_{A_+ B}$ at different orientations after the pump pulse.
One can see that $\rho_{A_+B}$ is antisymmetric about 90$^\circ$ and highly anisotropic,  such that according to Eq.~\eqref{eq:lambda},  the coherence signal between the $A_\pm$ and $B$ state is non-vanishing.
The bottom panel in Fig.~\ref{fig:rhoab} displays the degree of coherence (DOC),  defined by $g_{A_+B} = |\rho_{A_+B}|/ \sqrt{\rho_{A_+A_+} \rho_{BB}}$.
The DOC takes values ranging from 0 to 0.35 at different orientations.
Therefore,  while the DOC serves a useful indicator for the modulation in OD for atoms~\cite{rohringer2009, goulielmakis2010},  it is less informative for molecules,  and an orientation averaged indicator should be used.

\begin{figure}[t]
\includegraphics[scale=0.49]{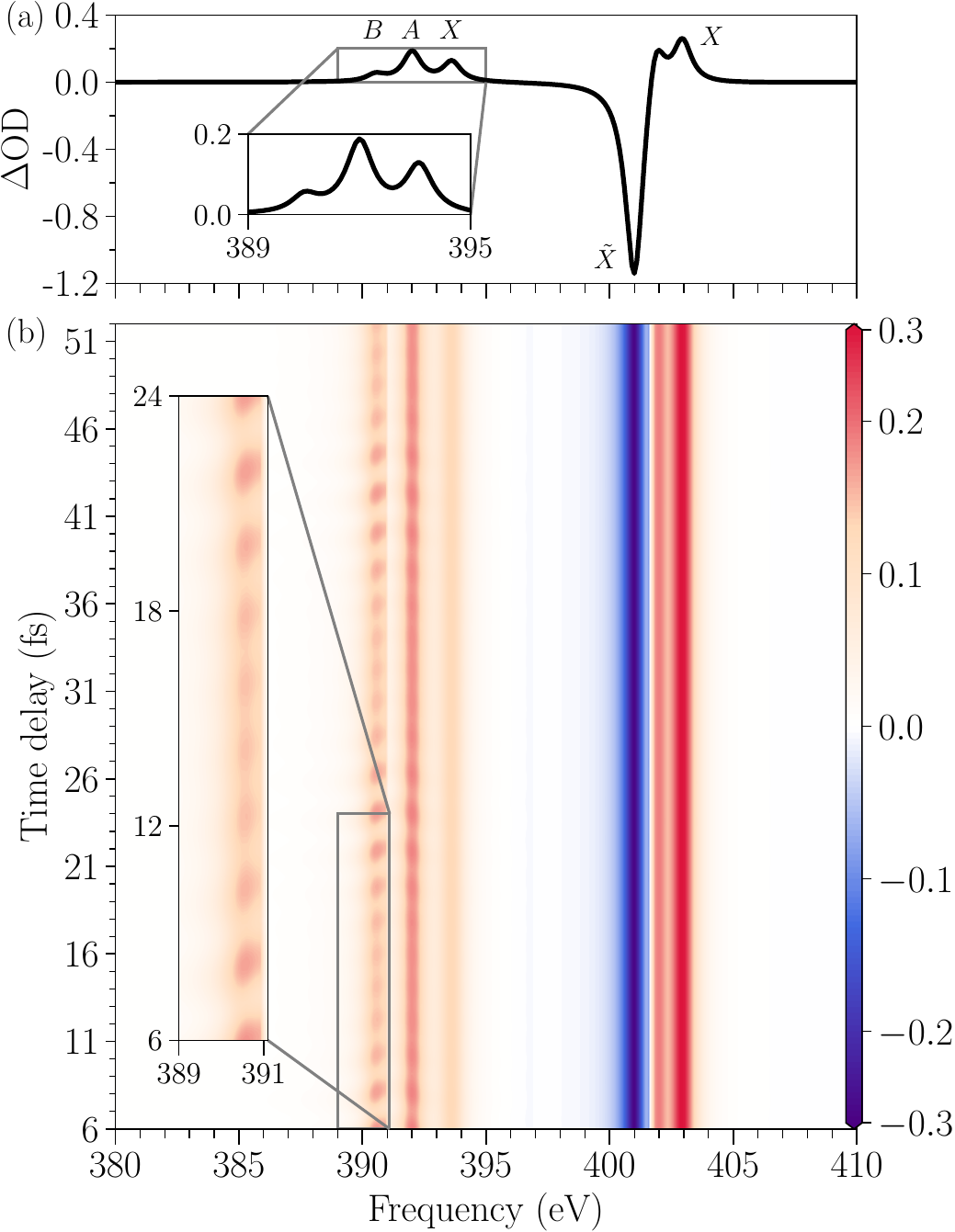}
\caption{(a) $\Delta$OD at 6 fs time delay for the $\tilde{X}$ state of N$_2$ and the $X,  A,$ and $B$ states of N$_2^+$.  The inset shows the spectrum between 389 to 395 eV.  (b) $\Delta$OD at different time delays.  The signal below 391 eV is enlarged by a factor of 3 and the signal between 397 to 401.5 eV is normalized to enhance the visual.  Modulation in the signal due to the $AB$ coherence is highlighted in the inset. }
\label{fig:ATAS}
\end{figure}

In the ATAS simulation,  we assume that the pump pulse and the probe pulse do not overlap,  such that the considered minimum time delay between the pulses is 6 fs.
The absorption spectrum ($\Delta$OD) at 6 fs time-delay is shown in Fig.~\ref{fig:ATAS}a.
The strong bleach of the signal at 401 eV is due to the depletion of the $\tilde{X}$ state,  where the orientation averaged population after the pump pulse is 0.7.
While the averaged population of the $X$,  $A$,  and $B$ states are all about 0.1,  the intensities of the absorption lines are different since for the $X$ to $1\sigma_u^{-1}$ , $A$ to $1\sigma_g^{-1}$,  and $B$ to $1\sigma_g^{-1}$ transition,  the relative ratios for the oscillator strength are $0.8: 1.0: 0.15$~\cite{glans1996} (see the inset of Fig.~\ref{fig:ATAS}a).

The orientation averaged absorption spectrum at different time delays is shown in Fig.~\ref{fig:ATAS}b.
In our modeling,  only the off-diagonal density matrix elements vary over time delay (see Eq.~\eqref{eq:FCF}).
As a result,  signals from the incoherent terms,  which come from the diagonal elements,  are constant over the time delays.
The absorption lines of the $\tilde{X}$ and $X$ states are therefore constant in time.
In contrast,  due to the contribution from the $\rho_{AB}$ term,  there are modulations in the signal from the $A$ and $B$ states.
But the modulation is much weaker for $A$ state than the $B$ state,
because the incoherent signal from the $A$ state is roughly 85\% larger than the $B$ state,  while the coherent signal is the same for both states.

\begin{figure}[ht]
\includegraphics[scale=0.4]{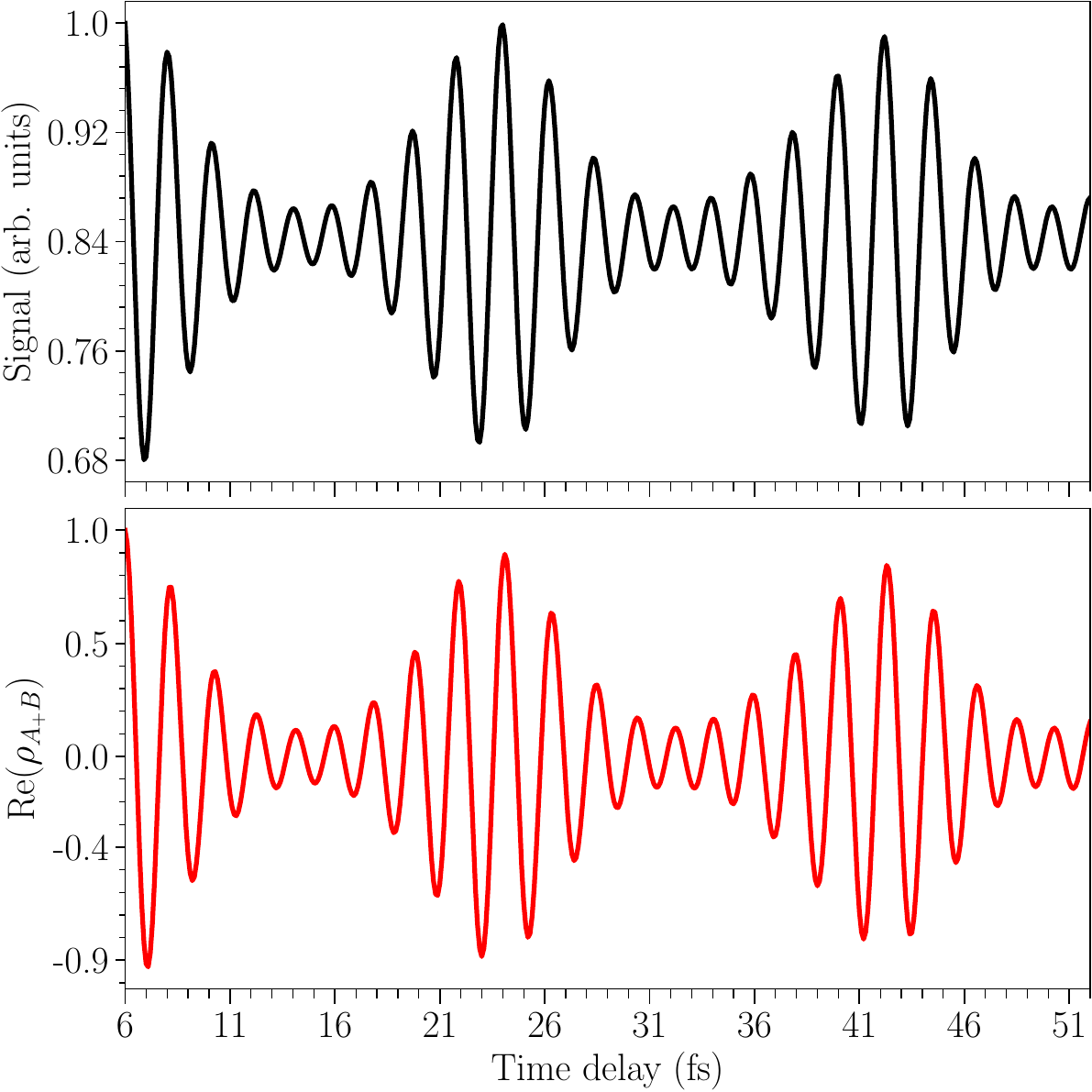}
\caption{Top: Integrated signal from 389 to 391 eV in Fig.~\ref{fig:ATAS}b.  Bottom: Real part of $\rho_{A_+B}$ at 45 degrees at different time delays.}
\label{fig:int_sig}
\end{figure}

The modulation in the signal from the $B$ state in the inset of Fig.~\ref{fig:ATAS}b exhibits clear dephasing and rephasing in a period of about 18 fs.
To compare the qualitative behavior between $\Delta$OD and vibronic coherence,  we plot the integrated signal in the 389 -- 391 eV region (orientation averaged) and the real part of $A_+B$ coherence at 45 degrees at different time delays in Fig.~\ref{fig:int_sig}.
This angle is chosen for the $A_+B$ coherence as it mimics the effect of orientation averaging.
We see that the integrated signal and the $A_+B$ coherence follow the same qualitative behavior.
This is expected since OD is proportional to the real part of the density matrix.
The maximum trough-to-peak ratio of the signal due to the varying vibronic coherence is about 0.68,  which should be distinguishable from the noise in experiments.

To this end,  we see that the contrast in the ATAS signal is related to the ratio between the coherent and incoherent signal.
Using Eqs.~\eqref{eq:chi} and \eqref{eq:lambda},  we can define a coherence contrast factor (CCF) for the absorption line of the $i$th state, 
\begin{align}
\mathrm{CCF}_i = \frac{ \sum_{\mu, \nu} d^\ast_{fj,  \nu}  d_{fi,  \mu} \mathrm{Re}(\lambda_{ij;\nu \mu} (t_{m}))}{\sum_{\mu, \nu} d^\ast_{fi,  \nu}  d_{fi,  \mu}  \mathrm{Re}(\lambda_{ii;\nu \mu})},
\label{eq:CCF}
\end{align}
where $t_{m}$ is the time delay when $\lambda_{ij;\nu \mu}$ is maximum and is taken as 6 fs here.
The trough-to-peak ratio of the signal is then given by $\rm (1-CCF)/(1+CCF)$.
If the CCF is small,  then the contrast in the signal due to the coherence will be weak.
For the absorption line of the $B$ state and $A$ state,  the CCF are 0.13 and 0.025,  respectively.
As a result,  the predicted trough-to-peak ratios are about 0.77 and 0.95 for the $B$ state and the $A$ state,
which agree with the qualitative behavior observed in Fig.~\ref{fig:ATAS}b.
The predicted ratio,  however,  is not expected to agree quantitatively with Fig.~\ref{fig:int_sig} due to the limited spectral resolution,  which causes the absorption line of the $A$ and $B$ state to overlap. 

In conclusion,  ATAS proves to be a robust method for probing vibronic coherence in charge migration. 
The necessary and sufficient conditions for ATAS to detect vibronic coherence between two arbitrary quantum states ($i$ and $j$) are as follows:

(i.)  Both states $i$ and $j$ must reach the same final state $f$ upon interaction with the attosecond XUV or x-ray pulse. 
This condition implies that,  in molecules with inversion symmetry,  ATAS can only probe coherence between states of the same parity.
In addition,  if states $i$ and $j$ have different electronic spin,  ATAS can detect their coherence only if there are spin-orbit couplings.

(ii.) The coherence contrast factor for either state $i$ or $j$ should be sufficiently large to distinguish the modulation due to changing vibronic coherence from the signal noise.
To verify whether this condition is satisfied,  acquiring the orientation-dependent density matrix of the pumped system and the transition dipole moments $\vec{d}_{fi}$ and $\vec{d}_{fj}$ is necessary.

With the robustness of ATAS being quantified,  we anticipate it to be an useful approach for studying vibronic coherence in molecules.
Another promising method with a solid theoretical foundation is the dissociative dication momentum spectroscopy (DDMS)~\cite{yuen2023b}.
The two methods could complement each other to provide complete characterization of vibronic coherence in a pumped molecule.
We eagerly anticipate future experimental validation of our predictions. 
With the necessary experimental techniques already well-established,  the full realization of monitoring electronic motion in generic molecules is within reach in the near future.


This work was supported by Chemical Sciences, Geosciences and Biosciences Division, Office of Basic Energy Sciences, Office of Science, U.S. Department of Energy under Grant No. DE-FG02-86ER13491.

%

\end{document}